\documentclass[aps,pra,twocolumn,showpacs,superscriptaddress,letterpaper]{revtex4}
\usepackage{amsmath,amssymb,graphicx,epsfig,color,bbm}
\usepackage[pass]{geometry}

\def\nn{\nonumber}

\begin{document}

\title{Quantum interference between independent reservoirs in open quantum systems}

\author{Ching-Kit Chan}
\affiliation{ITAMP, Harvard-Smithsonian Center for Astrophysics, Cambridge, MA 02138, USA}
\affiliation{Department of Physics, Harvard University, Cambridge, MA 02138, USA}
\author{Guin-Dar Lin}
\affiliation{ITAMP, Harvard-Smithsonian Center for Astrophysics, Cambridge, MA 02138, USA}
\affiliation{Department of Physics, Harvard University, Cambridge, MA 02138, USA}
\affiliation{Department of Physics, National Taiwan University, Taipei 10617, Taiwan}
\author{Susanne F. Yelin}
\affiliation{ITAMP, Harvard-Smithsonian Center for Astrophysics, Cambridge, MA 02138, USA}
\affiliation{Department of Physics, Harvard University, Cambridge, MA 02138, USA}
\affiliation{Department of Physics, University of Connecticut, Storrs, CT 06269, USA}
\author{Mikhail D. Lukin}
\affiliation{Department of Physics, Harvard University, Cambridge, MA 02138, USA}
\date{\today}

\begin{abstract}
When a quantum system interacts with multiple reservoirs, the environmental effects are usually treated in an additive manner. We show that this assumption breaks down for non-Markovian environments that have finite memory times. Specifically, we demonstrate that quantum interferences between independent environments can qualitatively modify the dynamics of the physical system. We illustrate this effect with a two level system coupled to two structured photonic reservoirs, discuss its origin using a non-equilibrium diagrammatic technique, and show an example when the application of this interference can result in an improved dark state preparation in a $\Lambda$ system.
\end{abstract}


\maketitle


\section{Introduction}

The study of environmental effects on a quantum mechanical system is at the heart of quantum information science. An environment generally results in quantum decoherence \cite{zurek03,schlosshauer05}. At the same time, it can be used to control the dynamics of a system through quantum reservoir engineering \cite{poyatos96,murch12,tomadin12,leghtas13,deffner13}, or generate interesting new phases in many body systems \cite{lee12,kessler12,fossfeig13,lin12,lee13}. The theoretical paradigm to study open quantum systems often relies on the master equation (ME) approach \cite{breuer02}. In practice, a system often couples to multiple reservoirs as is the case of cavity quantum electrodynamics (CQED) \cite{gea-banacloche05,armen06,majumdar12}, Jaynes-Cummings lattices \cite{nunnenkamp11}, photon-ion interfaces \cite{lamata11}, ion chain systems \cite{lin11}, phonon-induced spin squeezing \cite{bennett13}, etc. While it is commonly believed that different baths are additive in the ME in the Markovian limit, this assumption is not always valid for the more general non-Markovian physical situations \cite{coish04,yao06,taylor07,madsen11,hoeppe12,breuer09,huelga12,chin12,barnes12,kennes13,bermudez13}.

In this paper, we demonstrate and explain the emergence of quantum interference between independent baths and the corresponding impact on the reduced system.   We show that independent baths can produce correlations that cannot be described by an additive ME, in both short and long time regimes, except if all the baths are Markovian. This can be understood in the framework of Keldysh non-equilibrium diagrammatic theory \cite{keldysh65}, in which an additive ME misses all diagrams that represent interferences among multiple baths. As an example, we show that one can achieve a better dark state preparation of a driven $\Lambda$ system by taking the bath interference effects into account.

Our paper is outlined as follows. In Section \ref{sect_TLS}, we first illustrate this interference effect using a specific example of a two level system (TLS) coupled to two independent and structured photonic environments. This is followed by a general theoretical description in Section \ref {sect_general}, that shows how the environmental interference emerges dynamically when an open quantum system interacts with multiple reservoirs. We will discuss both the diagrammatic approach and the projection operator ME technique. In Section \ref{sect_3LS}, we present the applicable example of dark state preparation. We summarize in Section \ref{sect_conclusion} and provide some technical details in the Appendix.


\section{An illustrative example: two  photonic reservoirs}\label{sect_TLS}

\subsection{Additive ME}
It is usually assumed that the dynamics of a system coupled to independent baths is governed by an additive ME. An additive ME for the reduced density matrix $\rho_s(t)$ of the system has the form:
\begin{eqnarray}
\dot\rho_s(t) = \mathcal{\hat L}_1[\rho_s]+\mathcal{\hat L}_2[\rho_s]+... ,
\label{eq_additive_ME}
\end{eqnarray}
where the superoperator $\mathcal{\hat L}_i$ describes the noise effect due to bath $i$. For a Markovian environment, $\mathcal{\hat L}$ is a standard time-independent Lindblad operator, while in the non-Markovian regime, $\mathcal{\hat L}$ gains time dependence and can be a time-integral of the memory kernel operator \cite{breuer02}.

\begin{figure}[t]
\begin{center}
\includegraphics[angle=0, width=1\columnwidth]{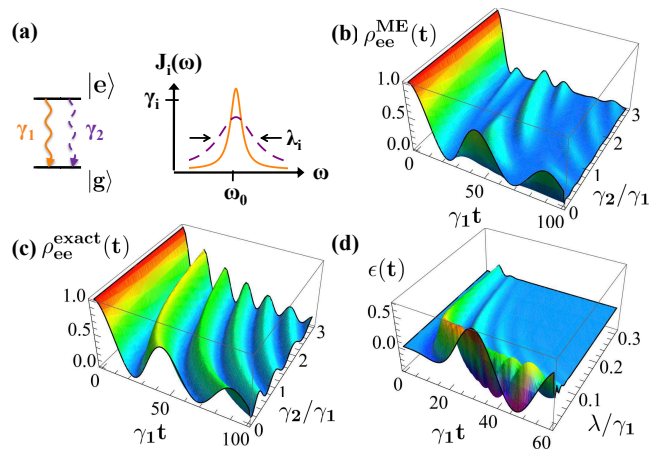}
\caption{(Color online) A demonstrative example showing the interference effect between two structured environments. (a) shows the model of a TLS interacting with two independent photonic vacua. Each photonic bath is characterized by a non-Markovian spectral density $J_i(\omega)$. (b) and (c) present the population dynamics of an initially excited TLS based on the additive ME and the exact solution, respectively. The ME solution produces a much suppressed oscillation due to the missing of the bath interference. The parameters used are $\lambda_1/\gamma_1 = 0.01$, $\lambda_2/\gamma_1=0.02$. Similar discrepancies are observed for other parameters. (d) plots the error $\epsilon(t)=\rho_s^{\text{exact}}(t)-\rho_s^{\text{ME}}(t)$ of the additive ME using $\lambda_1=\lambda_2=\lambda$ and $\gamma_1=\gamma_2$. As $\lambda$ increases, memory times are reduced and the additive ME result starts to approach the exact solution in the Markovian limit.}
\label{fig_twobaths}
\end{center}
\end{figure}

The issue of an additive ME can be best illustrated by a simple theoretical model: a two level system interacting with two uncoupled photonic reservoirs as shown in Fig.~\ref{fig_twobaths}(a). The canonical interaction is given by
\begin{align}
H &= \frac{\omega_0}{2}\sigma_z +\sum_{i=1,2}\sum_{k} \omega_k b_{i,k}^\dagger b_{i,k} + \sum_{i=1,2} V_i \nn \\
V_i &=  \sum_{k} g_{i,k} \left( b_{i,k}^\dagger \sigma_- + b_{i,k} \sigma_+ \right),
\end{align}
where $g_{i,k}$ is the coupling constant, $b_{i,k}^\dagger$ creates a photon in bath $i$, $\sigma_- = |g\rangle \langle e|$ and $\omega_0$ is the TLS energy splitting. The photonic environments are independent, so that $[b_{1,k},b_{2,k'}^\dagger]=0$. When the baths are initially in the vacuum state, the dynamics of the TLS is determined by the spectral densities defined as $J_i(\omega)= \int \frac{dt}{2\pi} \sum_k g_{i,k}^2 e^{i(\omega -\omega_k)t}$ \cite{gardiner00}. A constant $J(\omega)$ corresponds to the Markovian limit and a non-Markovian spectral density in general has a frequency dependence, e.g. the dynamics of a CQED system can be modeled by a Lorentzian $J(\omega)$. In the case of a single photonic bath (say $g_{1,k} \neq 0$, but $g_{2,k}=0$), this model can be solved using an exact ME that is applicable in both Markovian and non-Markovian regions \cite{vacchini10}.

If we intuitively assume that the presence of a second independent photonic bath just contributes an additional term to the ME, the additive ME will take the form:
\begin{eqnarray}
\dot\rho_s(t) = \left(\Gamma_1(t)+ \Gamma_2(t) \right) \Big[  \sigma_-  \rho_s(t) \sigma_+ - \frac{1}{2}\left\{ \sigma_+ \sigma_-, \rho_s(t)\right\} \Big],
\label{eq_ME_twobaths}
\end{eqnarray}
where the time dependent decay rate $\Gamma_i(t)$ due to bath $i$ originates from the non-Markovian properties of the baths \cite{breuer02,vacchini10}. We note that a non-Markovian ME in general has a integro-differential structure. The single bath case here is a specific example where the ME can take a time-convolutionless form \cite{breuer02}.

Fig.~\ref{fig_twobaths}(b) shows the evolution of the excited state population $\rho_{ee}(t)$ of an initially excited TLS in the presence of two independent photonic baths evaluated by this additive ME (Eq.~(\ref{eq_ME_twobaths})). We consider Lorentzian spectral densities $J_i(\omega) = \gamma_i \lambda_i^2 / \{2\pi [(\omega_0-\omega)^2 + \lambda_i^2]\}$, where $\lambda_i$ can be interpreted as the inverse memory time of the photonic bath $i$. This model can be mapped to the CQED, where $\lambda_i$ and $\sqrt{\gamma_i \lambda_i /2}$ play the roles of cavity decay rate and vacuum Rabi frequency, respectively. Using this spectral density the time dependent decay rate becomes $\Gamma_i(t) = 2\gamma_i\lambda_i \sinh(d_i t/2)/[d_i \cosh(d_i t/2)+\lambda_i \sinh(d_i t/2)]$, where $d_i=\sqrt{\lambda_i^2 -2 \gamma_i \lambda_i}$ \cite{breuer02,vacchini10}. In the absence of the second bath ($\gamma_2 = 0$), the TLS displays the expected damped vacuum Rabi oscillation. This correct behavior attributes to the exact and time-dependent $\Gamma_i(t)$ in the ME when there is only one bath. When we add another photonic bath ($\gamma_2 \neq 0$), the additive ME solution leads to a severely suppressed dynamics of the TLS.

The reduction of the coherence of the additive ME result is an artifact of the additive assumption. This can be revealed by the exact solution of the Schr$\ddot{\text{o}}$dinger equation of this model system that includes the entanglement dynamics between the system and the environments. The total system-environment wavefunction reads
\begin{eqnarray}
|\psi_{SE}(t)\rangle &=&c_e(t)|e\rangle|0,0\rangle+\sum_k c_{1,k}(t)|g\rangle|1_k,0\rangle \nn \\
 &&+ \sum_k c_{2,k}(t)|g\rangle|0,1_k\rangle + d |g\rangle|0,0\rangle ,
\label{eq_TLS_wavefunction}
\end{eqnarray}
where $|1_k,0\rangle$ denotes the state of a photon in bath $1$ and vacuum in bath $2$. Fig.~\ref{fig_twobaths}(c) shows the exact result using the same physical parameters. Apparently, the coherence of the vacuum Rabi oscillation is maintained despite the existence of an extra bath. For baths with similar spectral widths $\lambda_i$, the addition a second bath simply modifies the vacuum Rabi frequency, rather than the decay rate. These non-Markovian coherent features are missing in the additive ME solution. Unlike previous works \cite{breuer99,ferraro09} that showed the failure of the additive ME based on a small parameter expansion of the superoperator, we emphasize that the non-Markovian decay rates in Eq.~(\ref{eq_ME_twobaths}) are non-perturbative and exact in the single bath situation.

It is interesting to note that the discrepancies between the additive ME and the exact solution start to fade when both baths become Markovian. Fig.~\ref{fig_twobaths}(d) gives the difference between these two results as a function of the spectral width $\lambda=\lambda_1=\lambda_2$. As $\lambda$ increases, the spectral densities are flattened and the memory times are shortened. The revival behavior of the TLS is replaced by an exponential decay, and in this Markovian limit, the system can be well-approximated by an additive ME.

\subsection{Non-additive ME}

\begin{figure}[t]
\begin{center}
\includegraphics[angle=0, width=1\columnwidth]{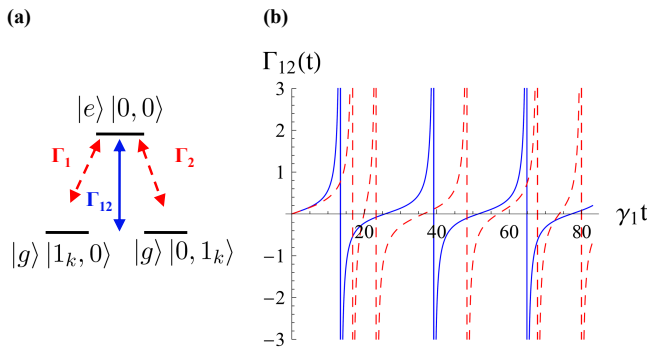}
\caption{(Color online) (a) shows the TLS oscillates between the excited and ground states with a non-Markovian decay rate $\Gamma_{12}(t)$. In the additive ME, the overall decay rate is approximated by an incoherent sum $\Gamma_{12}^{\text{exact}}(t) \approx \Gamma_{12}^{\text{additive}}(t) = \Gamma_1(t)+\Gamma_2(t)$. (b) shows the difference between $\Gamma_{12}^{\text{exact}}(t)$ (solid blue) and $\Gamma_{12}^{\text{additive}}(t)$ (dashed red). Here $\gamma_2=\gamma_1$, $\lambda_1/\gamma_1 = 0.01$, $\lambda_2/\gamma_1=0.02$.}
\label{fig_decay}
\end{center}
\end{figure}

Within this model, we have just demonstrated the deviation of the additive ME solution that assumes no bath interference from the exact solution that contains the full system-bath entanglement. We now show how the exact result can lead to a non-additive ME that contains explicit interference terms between the two bath parameters. Recalling the total system-environment wavefunction in Eq.~(\ref{eq_TLS_wavefunction}), one can show that the coefficients are governed by the following exact differential equations:
\begin{align}
\dot c_e(t) &= -\int_0^t dt' c_e(t') \left[ k_1(t-t') +k_2(t-t')\right] \nn \\
\dot c_{i,k}(t) &= -i g_{i,k}\int_0^t dt' c_e(t') e^{-i(\omega_0 - \omega_k) t'},
\label{eq_supple_ce}
\end{align}
where $k_i(t) = \sum_k g_{i,k}^2 e^{i(\omega_0-\omega_k)t}$ gives the memory kernel due to bath $i$ and is related to the spectral density through $k_i(t) = \int d\omega e^{i(\omega_0-\omega)t} J_i(\omega)$. Following the procedures in Ref. \cite{breuer99,breuer02,vacchini10}, one can obtain a ME by tracing out the bath degrees of freedom from the total entangled wavefunction. Considering $c_e(0)$ to be real, an exact and non-additive ME for this model system is:
\begin{align}
\dot\rho_s(t) =  \Gamma_{12}^{\text{exact}}(t) \Big[  \sigma_-  \rho_s^{\text{exact}}(t) \sigma_+ - \frac{1}{2}\left\{ \sigma_+ \sigma_-, \rho_s^{\text{exact}}(t)\right\} \Big],
\end{align}
where the non-Markovian and time-dependent decay rate $\Gamma_{12}^{\text{exact}}(t) = -2 \dot c_e(t) /c_e(t)$.

Using the Lorentzian spectral densities $J_i(\omega)$ used in the previous subsection, the memory kernels become $k_i(t) = \gamma_i \lambda_i e^{-\lambda_i |t|} /2$ and Eq.~(\ref{eq_supple_ce}) can be solved analytically. The resultant decay rate becomes:
\begin{align}
&\Gamma_{12}^{\text{exact}}(t) \nn \\
&= \frac{-2 \sum_i s_i e^{s_i t} (s_i+\lambda_1)(s_i+\lambda_2) \prod_{j\ \neq i, k \neq j} (s_j-s_k)}{\sum_i e^{s_i t} (s_i+\lambda_1)(s_i+\lambda_2) \prod_{j\ \neq i, k \neq j} (s_j-s_k)},
\label{eq_supple_rate}
\end{align}
where $s_i$ are the three roots of the equation $s(s+\lambda_1)(s+\lambda_2)+\gamma_1 \lambda_1 (s+\lambda_2)/2 +\gamma_2 \lambda_2 (s+\lambda_1)/2 = 0$. This exact decay rate clearly displays a mixing of physical parameters from the two baths. When the decay rate of either one of the baths vanishes, $\Gamma_{12}^{\text{exact}}$ reduces to the single bath situation as we described before:
\begin{align}
\Gamma_{12}^{\text{exact}}(t)  \xrightarrow{\gamma_{2}\rightarrow 0} \Gamma_{1} (t)  = \frac{2\gamma_{1}\lambda_{1} \sinh(d_{1} t/2)}{d_{1} \cosh(d_{1} t/2)+\lambda_{1} \sinh(d_{1} t/2)},
\end{align}
where $d_1=\sqrt{\lambda_1^2 -2 \gamma_1 \lambda_1}$ and a similar expression for $\Gamma_2(t)$ can be obtained in the same way. Now, we can see that the additive ME (i.e. Eq.~(\ref{eq_ME_twobaths})) approximates the overall decay rate by a sum of the decay rates of the two baths and thus misses the interference effect:
\begin{align}
\Gamma_{12}^{\text{additive}}(t) = \Gamma_1(t) + \Gamma_2(t) \neq \Gamma_{12}^{\text{exact}}(t).
\end{align}

Fig.~\ref{fig_decay}(b) shows the difference between $\Gamma^{\text{exact}}_{12}(t)$ and $\Gamma^{\text{additive}}_{12}(t)$. Since the density matrix is oscillating, the non-Markovian decay rate can in general be negative and divergent. Here, $\Gamma^{\text{exact}}_{12}(t)$ is periodic in time, implying a coherent oscillation between the excited state and the ground states. On the other hand, $\Gamma^{\text{additive}}_{12}(t)$ loses this feature because it originates from a sum of incoherent decays as shown in Fig.~\ref{fig_decay}(a).


\section{Quantum interference between two general reservoirs}\label{sect_general}

\subsection{Diagrammatic perspective}

In order to trace the origin of the issue of the additive ME, a microscopic theory that possesses the entanglement between the system and the environment is required. Here, we adopt the diagrammatic approach based on the Keldysh non-equilibrium Green's function technique \cite{keldysh65,saikin07,chan11,chan12}. Different from the conventional ME approach that traces away the environmental degree of freedom, the diagrammatic technique has the advantage that maintains the full dynamics of the entangled system-environment wavefunction. For a general system interacting with two environments through $V(t)=V_1(t)+V_2(t)$, the dynamics of a system observable $\mathcal O_S$ can be formally expressed as:
\begin{eqnarray}
\left\langle \mathcal O_S (t) \right \rangle&=& \left\langle \psi_{SE}(0)|\mathcal U^\dagger (t) \mathcal O_S \otimes \mathcal I_E \mathcal U(t) |\psi_{SE}(0) \right\rangle, \nn \\
\mathcal U(t)&=& \mathcal{T} \text{exp}\Big\{ -i\int_0^t dt'\left[ V_1(t')+V_2(t')\right] \Big\},
\label{eq_diagram}
\end{eqnarray}
where $\left| \psi_{SE}(0)\right\rangle$ is the initial system-environment state and the generalization to multiple reservoirs is straightforward. A diagrammatic expansion of Eq.~(\ref{eq_diagram}) allows the visualization of the underlying physical processes and the justification of the additive ME.

\begin{figure}[t]
\begin{center}
\includegraphics[angle=0, width=1\columnwidth]{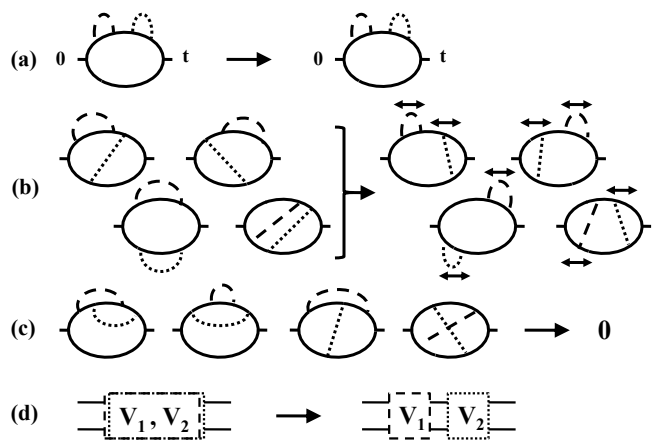}
\caption{Diagrammatic representations of the dynamics of a system interacting with two general baths from time $0$ to $t$. (a-c) Lowest order diagrams that describe the interference effect between bath $1$ (dashed lines) and $2$ (dotted lines). Left: exact diagrams; right: diagrammatic correspondence of the additive ME. Similar diagrams obtained by exchanging the bath lines and flipping the diagrams are not shown. All the diagrams have the same order. The additive ME solution does not allow the baths to have time overlaps in (b), nor to cross in (c). (d) shows that the additive ME corresponds to a ladder approximation in the diagrammatic structure and thus omits the interference between baths.}
\label{fig_diagram}
\end{center}
\end{figure}

To understand the correlated decoherence effect of the two baths, we depict the lowest order interference diagrams in Fig.~\ref{fig_diagram}(a-c). These diagrams are extracted from the expression:
\begin{eqnarray}
\int_0^{t} D^{(4)} t  \left\langle  \bigg[V(t_1),\Big[V(t_2),\big[V(t_3),[V(t_4),\mathcal O_S ]\big]\Big]\bigg]  \right\rangle,
\label{eq_4th_order}
\end{eqnarray}
where $\int_0^t D^{(n)} t = \int_0^t dt_n \int_0^{t_n}dt_{n-1} ... \int_0^{t_2} dt_1$. In order to simplify the physical picture, we have considered a separable initial state $ \left|\psi_{SE}(0) \right\rangle  =  \left|\psi_{S}(0) \right\rangle \otimes \left|\psi_{E_1}(0) \right\rangle \otimes \left|\psi_{E_2}(0) \right\rangle $, and assumed $\langle \psi_{E_i}(0)| V_i |\psi_{E_i}(0)\rangle = 0 $. Each diagram represents how the physical state $\left| \psi_{SE}\right\rangle$ evolves from the time $0$ (left) to $t$ (right) while undergoing system-bath interactions (dashed/dotted lines). It is clear that the two baths, though being independent, can affect the quantum system with finite time overlaps. The problem of the additive ME is ascribed to the missing of this kind of correlated influence from both baths.

For comparison purposes, we can perform a similar diagrammatic expansion of the additive ME governed by Eq.~(\ref{eq_additive_ME}). This can be achieved by finding the perturbative solution of the reduced density matrix from the additive ME using $V t$ as a small parameter. The lowest order results are shown in the right hand side of Fig.~\ref{fig_diagram}(a-c). We find that while the additive ME retains the non-overlapping diagram (Fig.~\ref{fig_diagram}(a)), it forbids the overlapping effect of the two baths (Fig.~\ref{fig_diagram}(b)) and excludes all crossing terms (Fig.~\ref{fig_diagram}(c)). In general, the additive ME assumption corresponds to a ladder-like diagrammatic structure as shown in Fig.~\ref{fig_diagram}(d). The missing diagrams are of the same order as the non-crossing counterparts and should be included to recover the true dynamics of the reduced physical system. This analysis can be easily extended to the higher orders. The inclusion of higher order terms in the expansion of each $ \mathcal {\hat L}_i[\rho_s]$ cannot compensate the missing interference diagrams.

It is known that the presence of an initial system-environment correlation can modify the ME \cite{breuer02,gessner11,roghani11,laine12}. We note, however, that our correlation effect emerges dynamically and is present even when the system and environments are initially separable.

The disappearance of this interference bath effect in the Markovian limit can be understood with the aid of diagrams as well. For two Markovian baths with zero memory times, i.e. $\left \langle V_{1(2)}(t_i) V_{1(2)}(t_j)\right\rangle \sim \delta (t_i-t_j)$, the two ends of each dashed (dotted) line would collapse to the same point in the diagram. In this regime, all the crossing diagrams vanish and the diagrams are reduced to have a ladder structure. The baths can no longer affect the system with time overlaps and the additive ME becomes a good approximation. We stress that this situation requires all the baths being Markovian. The correlation effect still exists in the case of one non-Markovian bath plus multiple Markovian reservoirs, as we will see in Section \ref{sect_3LS}.

\subsection{Projection operator ME framework}

Another theoretical framework for open quantum system with non-Markovian environments can be carried out using the Nakajima-Zwanzig projection operator technique \cite{breuer99}. This would help us to have a parallel understanding of how the bath interference terms arise in the corresponding non-additive ME, compared to the diagrammatic treatment above.

The idea of this approach is to start with an exact Liouville equation that describes the dynamics of the total system-environment density matrix $\rho_{SE}(t)$, then project the state on relevant and irrelevant parts, and trace out the environment at the end to obtain a formal ME. The derivation procedure is detailed in Ref.~\cite{breuer02} and we shall not repeat it here. Defining $L  \rho = -i [V , \rho]  $ and  the projection operator $P\rho_{SE}(t)  = \text{Tr}_E \left \{ \rho_{SE}(t) \right\} \otimes \rho_E(0)$, the resultant ME is given by:
\begin{eqnarray}
\frac{\partial}{\partial t} P \rho_{SE}(t) = \int_0^t dt'   \hat K(t,t') P \rho_{SE} (t')
\label{eq_projectME}
\end{eqnarray}
where we again assumed a separable initial state and $\text{Tr}_E \{ V\} = PLP =0$. The memory kernel superoperator $\hat K(t,t')$ can be expressed as:
\begin{eqnarray}
\hat K(t,t') &=& P \hat L(t) \hat G(t,t') (1-P) \hat L(t') P, \nn \\
\hat G(t,t') &=& \mathcal{T}  \text{exp} \left\{    \int_{t'}^t  dt'' (1-P) \hat L(s'')\right\}.
\end{eqnarray}
Eq.~(\ref{eq_projectME}) is formally exact, but hard to implement in practice due to the complicated memory kernel. Yet, we can still observe how the bath interference terms appear from this formal ME treatment. Taking $V=V_1+V_2$ and expand Eq.~({\ref{eq_projectME}}) in the powers of $V$, we find an non-additive ME:
\begin{widetext}
\begin{eqnarray}
\dot \rho_{s}(t) &=& -\sum_{i=1,2} \int_0^t dt_1 \text{Tr}_E \left \{   [V_i(t) , [V_i(t_1), \rho_{s} (t_1) ]]\right\} \nn  \\
&&-\int_0^t D^{(3)}t \text{Tr}_E  \left\{   [V(t),[V(t_3),[V(t_2),[V(t_1),\rho_s(t_1)]]]] -  [V(t),[V(t_3), \text{Tr}_E \left\{[V(t_2),[V(t_1),\rho_s(t_1)]]  \right\} ]]  \right\} ,\nn \\
&&+ O(V^6)
\label{eq_projectME_expand}
\end{eqnarray}
\end{widetext}
The first terms correspond to the Born (without Markov) approximation. The second terms include interference ($\sim V_1^2 V_2^2$) and non-interference ($\sim V_1^4 \ \text{or}\  V_2^4$) contributions. These interference terms share the same physical origin as those in Eq.~(\ref{eq_4th_order}) obtained from the diagrammatic calculation.

When both environments are Markovian, i.e. $\text {Tr}_E \{ V_1 (t)V_1 (t')\} \sim \text {Tr}_E \{ V_2 (t)V_2 (t')\} \sim \delta(t-t')$, it is straightforward to show that the interference terms would vanish; while when one of the baths is non-Markovian, the interference remains, in agreement with the diagrammatic result. In general,  it is difficult to solve Eq.~({\ref{eq_projectME_expand}}) due to its multi-integro-differential structure. It would be more direct to calculate the perturbative solution as we will see in the example of the following section.

\begin{figure}[t]
\begin{center}
\includegraphics[angle=0, width=1\columnwidth]{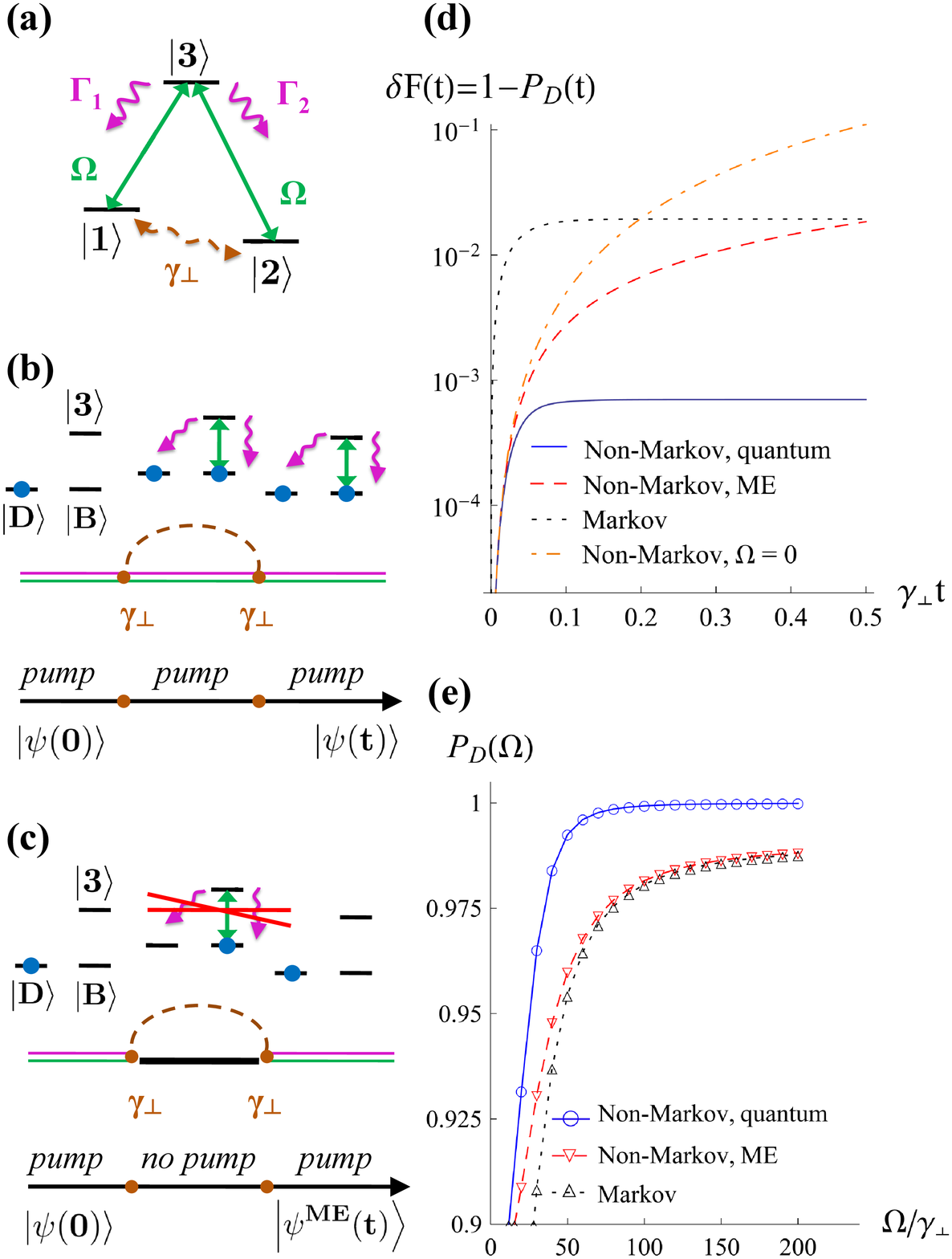}
\caption{(Color online) Dephasing comparison of a driven $\Lambda$ system. (a) The model system in the presence of Markovian relaxations and non-Markovian dephasing. (b) shows the virtual dephasing evolution, in which the initial dark state $|D\rangle= (|1\rangle -|2\rangle)/\sqrt 2$ is flipped to the bright state and then pumped back. (c) represents the corresponding evolution of the additive ME solution. The missing of the bath interference prohibits the simultaneous action of the pumping (double line) and dephasing processes (dashed line). (d) compares the dark state infidelity $\delta F(t)=1-P_D (t)$ calculated by various approaches using $\Omega=\Gamma_1=\Gamma_2=100~\gamma_\perp$. (e) provides the dark state population reached by increasing the driving field when $\gamma_\perp t= 1/2$ and $\Gamma_1=\Gamma_2=100~\gamma_\perp$. The additive ME predicts a lower dark state fidelity.}
\label{fig_3LS}
\end{center}
\end{figure}


\section{Applications}\label{sect_3LS}

\subsection{Dephasing of a $\Lambda$ system}

We have just shown how the bath interference can affect the dynamics of the reduced system and how it emerges dynamically due to the system-environment entanglement. We now turn to an explicit example in which the inclusion of interferences between independent baths could yield a reduced decoherence of the physical system. Consider a driven three level $\Lambda$ system under the influence of Markovian atomic relaxations and non-Markovian dephasing as shown in Fig.~\ref{fig_3LS}(a). The resonantly driven Hamiltonian and the system-bath interactions are given by:
\begin{eqnarray}
H_\Omega &=& \frac{\Omega}{2}\left( \sigma_{13}+\sigma_{23} + h.c.\right),\nn \\
V_1 &=& \sum_k \left( g_{1k} \sigma_{31} a_{1k} +g_{2k} \sigma_{32} a_{2k}  +h.c.\right), \nn \\
V_2 &=& \sigma_z \sum_k \lambda_k \left( b_k + b_k^\dagger \right),
\end{eqnarray}
where $\sigma_{ij}=|i\rangle\langle j|$, $\sigma_z = \sigma_{22}-\sigma_{11}$, and $a_k$ and $b_k$ are bosonic bath operators. The electric-dipole couplings $g_{i,k}$ are responsible for the corresponding population relaxations, and we adopt a bosonic dephasing model \cite{lidar12} with coupling constant $\lambda_k$. In the dressed state picture, the optical drive $\Omega$ together with the atomic decays $\Gamma_i$ pump the system to the dark state $|D\rangle= (|1\rangle -|2\rangle)/\sqrt 2$; meanwhile, the dephasing term would decohere it to a classical mixture of dark and bright states $|B\rangle= (|1\rangle +|2\rangle)/\sqrt 2$. In the following, we shall examine the dark state fidelity as a consequence of these two competing dissipative processes with and without the presence of bath interferences.

The Markovian relaxation and non-Markovian dephasing rates are defined through:
\begin{eqnarray}
\langle V_1(t) V_1(t')\rangle &=&  \left( \Gamma_1 +\Gamma_2 \right)  \delta(t-t')\sigma_{33}, \nn \\
\langle V_2(t) V_2(t')\rangle &= & \frac{\gamma_\perp^2}{2} K(t-t') \sigma_z^2,
\end{eqnarray}
where $K(\tau)$ is the dephasing memory function. To the second order of $\gamma_\perp$, an additive ME for this model is given by:
\begin{eqnarray}
\dot\rho_s(t) &=&-i[H_\Omega,\rho_s(t)] - \sum_{i=1,2} \frac{\Gamma_i}{2}\mathcal{D}[\sigma_{i3}]\rho_s(t)\nn \\
&&\ -\frac{\gamma_\perp^2}{2} \int_0^t dt' K(t-t') \mathcal{D}[\sigma_{z}]\rho_s(t'),
\label{eq_3LS_ME}
\end{eqnarray}
where $\mathcal{D}[\mathcal O]\rho_s = \{\mathcal  O ^\dagger\mathcal O, \rho_s\}-2 \mathcal O\rho_s \mathcal O^\dagger$. On the other hand, we can obtain the full equation of motions for the system and bath operators. Upon eliminating the bath operators, we arrive at an exact and reduced Heisenberg equation of motion:
\begin{eqnarray}
\label{eq_3LS_Heisenberg}
&&\langle \dot \sigma_{ij}(t)\rangle = i\langle[ H_\Omega(t) ,\sigma_{ij}(t)]\rangle - \sum_{i=1,2} \frac{\Gamma_i}{2}\left\langle \mathcal{D}'[\sigma_{3i}(t)] \sigma_{ij}(t)\right\rangle \nn \\
&&\ \ \ \ \ \  -\frac{\gamma_\perp^2}{2} \int_0^t dt' K(t-t')\left\langle  \big[\sigma_z(t')   ,\left[\sigma_z(t)   , \sigma_{ij}(t) \right] \big]\right\rangle,
\end{eqnarray}
where $\mathcal{D}'[\mathcal O]\sigma = \{\mathcal  O \mathcal O ^\dagger, \sigma\}-2 \mathcal O\sigma \mathcal O^\dagger$. Unlike Eq.~(\ref{eq_3LS_ME}), the presence of the two time correlation functions in this exact quantum equation allows the interference between the relaxations and the dephasing processes. These two approaches are generally different except in the Markovian dephasing limit such that $K(\tau)=\delta(\tau)/\gamma_\perp$. In the following, we are going to compare the additive ME (Eq.~(\ref{eq_3LS_ME})) and the quantum solution (Eq.~(\ref{eq_3LS_Heisenberg})). We solve Eq.~(\ref{eq_3LS_Heisenberg}) perturbatively up to $O(\gamma_\perp^2)$ and compare it with the solution of Eq.~(\ref{eq_3LS_ME}), which is valid to the same order of $\gamma_\perp$. This corresponds to a perturbative solution of the problem that is second order in the non-Markovian dephasing interaction ($V_2$), but infinite order in the Markovian relaxation interaction ($V_1$).

Fig.~\ref{fig_3LS}(d) and (e) provide the time and field dependences of the dark state population computed by solving the Heisenberg equations (quantum), the additive ME and the case of Markovian dephasing. We consider a long memory time such that $K(\tau)\approx 1$ for the non-Markovian calculations in the time regime of interest. The quantum solution demonstrates a much suppressed dephasing than that of the additive ME, which behaves similarly to the Markovian result. For $\Omega \gg \Gamma_1 \approx \Gamma_2$, the dark state fidelities are $\sim O[(\gamma_\perp / \Gamma_1)^2]$ and $\sim O[\gamma_\perp / \Gamma_1]$ for the quantum and additive ME solutions, respectively. Together with the TLS example in Fig.~\ref{fig_twobaths}, the result of this dephasing problem indicate the inadequacy of the additive ME in both the long and short time scales.

The diagrammatic representation offers a visualization of the difference of the underlying mechanisms between these two approaches. Fig.~\ref{fig_3LS}(b) shows the virtual dark state evolution. To the second order of $\gamma_\perp$, the dephasing interaction flips the initial state from $|D\rangle$ to $|B\rangle$, which is then pumped back to $|D\rangle$, and this procedure repeats the second time. In this picture, the relaxations and dephasing can act simultaneously on the system. However, in the additive ME that neglects such an interference, the pumping is forbidden during the virtual dephasing process as presented in Fig.~\ref{fig_3LS}(c). This discrepancy eventually leads to a reduced dark state fidelity of the additive ME solution. A more detailed diagrammatic representation based on the reduced density matrix and another additive ME that includes the dressing effect of the drive can be found in the Appendix.

\subsection{Applicability for general quantum systems}

The dephasing problem above provides a particular scenario where the interference between environments is important and can help the reduction of decoherence. It is difficult to carry out the full ME, like Eq.~(\ref{eq_projectME_expand}), in order to analyze such an effect given this relatively simple example. For general and more complex physical circumstances, the bath interference can either assist or deteriorate the coherence of the physical system. The diagrammatic procedure can be straightforwardly applied to study this interference effect in open quantum systems that involve at least one non-Markovian bath. This would include many body systems or problems with multi-level structures, where an exact solution is not available and it is difficult to carry out a full ME simulation. For simple few-level systems coupled to Markovian baths only, the bath interference is negligible and the conventional additive ME approach would be a simpler theoretical tool to calculate the corresponding decoherence dynamics.


\section{Conclusions}\label{sect_conclusion}

We show that the additive assumption of independent environments in the ME is valid only when all the baths are Markovian. Because of the dynamical entanglement between the system and the environment, quantum interferences between multiple environments can emerge and strongly alter the decoherence mechanism for non-Markovian circumstances. A diagrammatic technique allows a direct justification of the additive assumption. We also provide a realistic example in which the inclusion of this interference effect could lead to a lower decoherence of the physical system. We believe the important role of the system-environment entanglement provides a new direction to understand and suppress decoherence for general open quantum systems.

\acknowledgements

We thank Tony Lee and Shenshen Wang for very helpful discussions. This work is supported by NSF, CUA, ITAMP, AFOSR and MURI. C.K.C. is supported by the Croucher Foundation.

\appendix

\begin{widetext}

\section{Alternative additive ME for the dephasing problem}

In the main text, we describe one possible additive ME for the driven dephasing $\Lambda$ system. The equation is rewritten here:
\begin{align}
\dot\rho_s(t) = -i[H_\Omega,\rho_s(t)] - \sum_{i=1,2} \frac{\Gamma_i}{2}\mathcal{D}[\sigma_{i3}]\rho_s(t) -\frac{\gamma_\perp^2}{2} \int_0^t dt' K(t-t') \mathcal{D}[\sigma_{z}]\rho_s(t').
\label{eq_supple_3LS_ME1}
\end{align}
Not only does this equation assume the relaxations ($\Gamma_i$) and dephasing ($\gamma_\perp$) processes are additive, but also treats the driving term in the same manner. An alternative way is to include the dressing effect of the optical drive on the dissipative processes. We will study this alternative in the following. To obtain this alternative additive ME, we work in the interaction picture (by treating $H_\Omega$ as the bare Hamiltonian) such that the dissipative operators are dressed by the driving Hamiltonian, and then perform the standard Born approximation for the dephasing interaction. The resultant additive ME (we denote as ME-2) is:
\begin{align}
\dot\rho_s(t) &= -i[H_\Omega,\rho_s(t)] - \sum_{i=1,2} \frac{\Gamma_i}{2}\mathcal{D}[\sigma_{i3}]\rho_s(t) \nn \\
&\ \  \ -\frac{\gamma_\perp^2}{2} \int_0^t dt' K(t-t') \left[ \sigma_z U(t-t') \sigma_z \rho_s(t') U(t'-t) - \sigma_z U(t-t') \rho_s(t') \sigma_z U(t'-t) + \text{h.c.} \right ],
\label{eq_supple_3LS_ME2}
\end{align}
where the driving propagator $U(t) = e^{-i H_\Omega t}$. Note that $U(t)$ only enters through the non-Markovian dephasing memory kernel rather than the Markovian relaxation terms, and this equation still contains no interference between the two dissipative processes.

Eq.~(\ref{eq_supple_3LS_ME1}) and (\ref{eq_supple_3LS_ME2}) can be solved directly by the Laplace transformation technique. Fig.~\ref{fig_supple_ME2} presents typical time evolutions obtained from these two additive ME approaches and the quantum method that include the bath interference (i.e. Eq.~(\ref{eq_3LS_Heisenberg})). We see that the ME-2 shows an unphysical result (Fig.~\ref{fig_supple_ME2}(c)) that it does not display the dephasing and it violates the positivity of the reduced density matrix. The violation of the positivity is not uncommon for non-Markovian MEs \cite{breuer02}. In this case, the failure is attributed to the fact that  Eq.~(\ref{eq_supple_3LS_ME2}) only includes the interference between the drive and the dephasing, but still omits that between the two dissipative processes. During the virtual dephasing process of the $\Lambda$ system, this equation allows the influence of the optical drive, but not the relaxation, resulting in an oscillatory dark state population. Therefore, this ME-2 approach still gives an incorrect dynamics of the reduced density matrix. The diagrammatic construction in the next section provides a better visualization of the underlying virtual processes.

\begin{figure}[t]
\begin{center}
\includegraphics[angle=0, width=0.9\columnwidth]{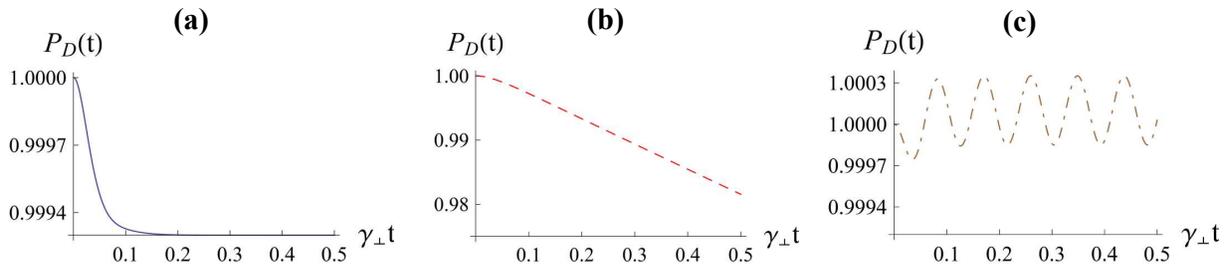}
\caption{(Color online) A comparison of the dark state populations of the driven $\Lambda$ system using various approaches. (a) the quantum solution, (b) the additive ME that neglects the bath interference, and (c) the additive ME-2 that includes the dressing of the optical drive, but still omits the bath interference. Parameters used: $\Omega=\Gamma_1=\Gamma_2=100~\gamma_\perp$.}
\label{fig_supple_ME2}
\end{center}
\end{figure}

\section{Diagrammatic comparison for the dephasing problem}

Here, we are going to give a detailed diagrammatic comparisons for the virtual dephasing processes of $\Lambda$ system problem based on the quantum solution, the additive ME, the additive ME-2 and the Markovian solution for the dephasing problem. In the regime of $\gamma_\perp t \ll 1$, the perturbative solution of the exact reduced Heisenberg equation of motion (Eq.~(\ref{eq_3LS_Heisenberg})) for the $\Lambda$ system is given by:
\begin{align}
\left\langle \mathcal \sigma_{ij} (t) \right \rangle = \left\langle \mathcal \sigma_{ij}^I(t)  \right \rangle -  \frac{\gamma_\perp^2}{2} \int_0^{t} dt_2 \int_0^{t_2} dt_1 K(t_2-t_1) \left\langle  \big[\sigma_z^I(t_1),[\sigma_z^I(t_2),\mathcal \sigma_{ij}^I (t)]\big]  \right\rangle +...,
\label{eq_3LS_expansion}
\end{align}
where we have used the interaction picture $\mathcal O^I(t) = e^{i (H_\Omega + V_1) t} \mathcal O e^{-i (H_\Omega + V_1) t}$. The second terms of Eq.~(\ref{eq_3LS_expansion}) is depicted in Fig.~\ref{fig_supple_diagramcomparison}(a), where we see that both the driving and the relaxation processes dress the reduced $\Lambda$ system and pump the system to the dark state $|D\rangle$. In the meanwhile, the system undergoes virtual dephasing process that flips the system between the dark and bright states.  This expression is second order in $\gamma_\perp$, but infinite order in $\Omega$, $\Gamma_1$ and $\Gamma_2$.

To make connection with the ME, the upper and lower panels in Fig.~\ref{fig_supple_diagramcomparison} correspond to the terms $\rho_s(t') \sigma_z(t') \sigma_z(t'')$ and $\sigma_z(t'') \rho_s(t') \sigma_z(t')$, respectively. The term $\sigma_z(t'') \sigma_z(t') \rho_s(t')$ can be obtained by flipping the upper diagram.

The corresponding diagrammatic representations for the perturbative solutions of the additive ME and ME-2 approaches (i.e. Eq.~(\ref{eq_supple_3LS_ME1}) and (\ref{eq_supple_3LS_ME2})) are presented in Fig.~\ref{fig_supple_diagramcomparison}(b-c). In these two approaches, the relaxation is forbidden during the virtual dephasing processes and thus the bath interference is neglected. They both lead to incorrect virtual states during the dephasing processes, regardless whether we include the dressing of the optical drive or not. On the other hand, the diagrams in the Markovian dephasing limit (i.e. $K(\tau) = \delta(\tau)/\gamma_\perp$) given in Fig.~\ref{fig_supple_diagramcomparison}(d) shows some resemblance to those of the additive ME without the virtual states in Fig.~\ref{fig_supple_diagramcomparison}(b). This explains why these two approaches have similar dark state populations as shown in Fig.~\ref{fig_3LS}(d-e).

\begin{figure}[t]
\begin{center}
\includegraphics[angle=0, width=0.9\columnwidth]{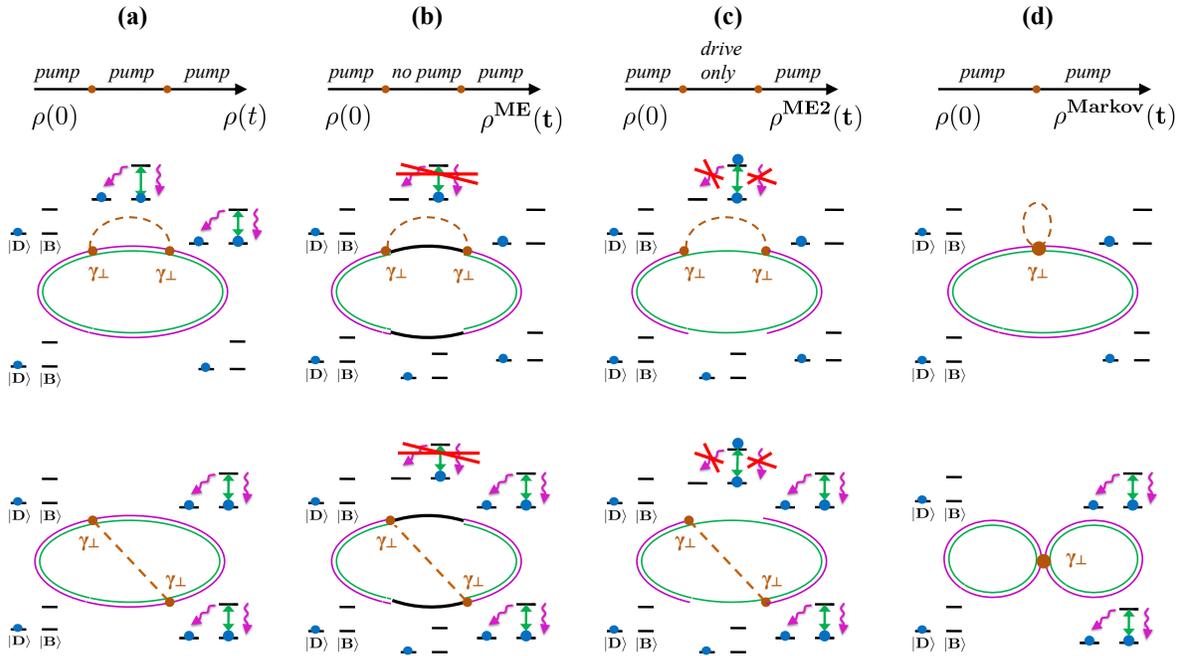}
\caption{(Color online) Diagrammatic constructions for the dephasing problem based on different theoretical techniques. Each diagram, being second order in the dephasing interaction, represents the virtual evolution of the system from time $0$ (left) to $t$ (right) with the dressing of the optical pumping (double purple/green) and the influence of the dephasing (dashed brown). The energy levels next to the diagram gives the virtual processes of the dark state evolution. (a) gives the quantum solution. (b) shows the additive ME approach that forbids both the relaxation and optical driving during the virtual dephasing process. (c) show the additive ME-2 result that permits the drive but not the relaxation during the virtual process. (d) presents the limit of Markovian dephasing. Similar diagrams obtained by flipping the bubbles are not shown.}
\label{fig_supple_diagramcomparison}
\end{center}
\end{figure}

\end{widetext}



\begin{thebibliography}{99}
\bibitem{zurek03} W. H. Zurek, Rev. Mod. Phys. \textbf{75}, 715 (2003).
\bibitem{schlosshauer05} M. Schlosshauer, Rev. Mod. Phys. \textbf{76}, 1267 (2005).
\bibitem{poyatos96} J. F. Poyatos, J. I. Cirac, and P. Zoller, Phys. Rev. Lett. \textbf{77}, 4728 (1996).
\bibitem{murch12} K.W. Murch, U. Vool, D. Zhou, S. J. Weber, S. M. Girvin, and I. Siddiqi, Phys. Rev. Lett. \textbf{109}, 183602 (2012).
\bibitem{tomadin12} A. Tomadin, S. Diehl, M. D. Lukin, P. Rabl, and P. Zoller, Phys. Rev. A \textbf{86}, 033821 (2012).
\bibitem{leghtas13} Z. Leghtas, U. Vool, S. Shankar, M. Hatridge, S. M. Girvin, M. H. Devoret, and M. Mirrahimi, Phys. Rev. A \textbf{88}, 023849 (2013).
\bibitem{deffner13} S. Deffner and E. Lutz, Phys. Rev. Lett. \textbf{111}, 010402 (2013).
\bibitem{lee12} T. E. Lee, H. H$\ddot{\text{a}}$ffner, and M. C. Cross, Phys. Rev. Lett. \textbf{108}, 023602 (2012).
\bibitem{kessler12} E. M. Kessler, G. Giedke, A. Imamoglu, S. F. Yelin, M. D. Lukin, and J. I. Cirac, Phys. Rev. A \textbf{86}, 012116 (2012).
\bibitem{fossfeig13} M. Foss-Feig, K. R. A. Hazzard, J. J. Bollinger, and A. M. Rey, Phys. Rev. A \textbf{87}, 042101 (2013).
\bibitem{lin12} G.-D. Lin and S. F. Yelin, Phys. Rev. A \textbf{85}, 033831 (2012).
\bibitem{lee13} T. E. Lee, S. Gopalakrishnan, and M. D. Lukin, Phys. Rev. Lett. \textbf{110}, 257204 (2013).
\bibitem{breuer02} H.-P. Breuer and F. Petruccione, \textit{Theory of Open Quantum Systems} (Oxford University Press, New York, 2002).

\bibitem{gea-banacloche05} J. Gea-Banacloche, T. C. Burt, P. R. Rice, and L. A. Orozco, Phys. Rev. Lett. \textbf{94}, 053603 (2005).
\bibitem{armen06} M. A. Armen and H. Mabuchi, Phys. Rev. A \textbf{73}, 063801 (2006).
\bibitem{majumdar12} A. Majumdar, D. Englund, M. Bajcsy, and J. Vu\v{c}kovi\'{c}, Phys. Rev. A \textbf{85}, 033802 (2012).
\bibitem{nunnenkamp11} A. Nunnenkamp, J. Koch, and S. M. Girvin, New J. Phys. \textbf{13}, 095008 (2011).
\bibitem{lamata11} L. Lamata, D. R. Leibrandt, I. L. Chuang, J. I. Cirac, M. D. Lukin, V. Vuleti\'{c}, and S. F. Yelin, Phys. Rev. Lett. \textbf{107}, 030501 (2011).
\bibitem{lin11} G.-D. Lin and L.-M. Duan, New J. Phys. \textbf{13}, 075015 (2011).
\bibitem{bennett13} S. D. Bennett, N.Y. Yao, J. Otterbach, P. Zoller, P. Rabl, and M. D. Lukin, Phys. Rev. Lett. \textbf{110}, 156402 (2013).


\bibitem{coish04} W. A. Coish and D. Loss, Phys. Rev. B \textbf{70}, 195340 (2004).
\bibitem{yao06} W. Yao, R.-B. Liu, and L. J. Sham, Phys. Rev. B \textbf{74}, 195301 (2006).
\bibitem{taylor07} J. M. Taylor, J. R. Petta, A. C. Johnson, A. Yacoby, C. M. Marcus, and M. D. Lukin, Phys. Rev. B \textbf{76}, 035315 (2007).
\bibitem{breuer09} H.-P. Breuer, E.-M. Laine, and J. Piilo, Phys. Rev. Lett. \textbf{103}, 210401 (2009).

\bibitem{madsen11} K. H. Madsen, S. Ates, T. Lund-Hansen, A. L$\ddot{\text{o}}$ffler, S. Reitzenstein, A. Forchel, and P. Lodahl, Phys. Rev. Lett. \textbf{106}, 233601 (2011).
\bibitem{hoeppe12} U. Hoeppe, C. Wolff, J. K$\ddot{\text{u}}$chenmeister, J. Niegemann, M. Drescher, H. Benner, and K. Busch, Phys. Rev. Lett. \textbf{108}, 043603 (2012).
\bibitem{huelga12} S. F. Huelga, $\acute{\text{A}}$. Rivas, and M. B. Plenio, Phys. Rev. Lett. \textbf{108}, 160402 (2012).
\bibitem{chin12} A. W. Chin, S. F. Huelga, and M. B. Plenio, Phys. Rev. Lett. \textbf{109}, 233601 (2012).
\bibitem{barnes12} E. Barnes, $\L$ukasz Cywi$\acute{\text{n}}$ski, and S. Das Sarma, Phys. Rev. Lett. \textbf{109}, 140403 (2012).
\bibitem{kennes13} D. M. Kennes, O. Kashuba, M. Pletyukhov, H. Schoeller, and V. Meden, Phys. Rev. Lett. \textbf{110}, 100405 (2013).
\bibitem{bermudez13} A. Bermudez, T. Schaetz, and M. B. Plenio, Phys. Rev. Lett. \textbf{110}, 110502 (2013).

\bibitem{keldysh65} L. V. Keldysh, JETP. \textbf{20}, 1018 (1965).
\bibitem{gardiner00} C. Gardiner and P. Zoller, \textit{Quantum Noise: A Handbook of Markovian and Non-Markovian Quantum Stochastic Methods with Applications to Quantum Optics} (Springer, New York, 2000).
\bibitem{vacchini10} B. Vacchini and H.-P. Breuer, Phys. Rev. A \textbf{81}, 042103 (2010).
\bibitem{breuer99} H.-P. Breuer, B. Kappler, and F. Petruccione, Phys. Rev. A \textbf{59}, 1633 (1999).
\bibitem{ferraro09} E. Ferraro, M. Scala, R. Migliore, and A. Napoli, Phys. Rev. A \textbf{80}, 042112 (2009).
\bibitem{saikin07} S. K. Saikin, W. Yao, and L. J. Sham, Phys. Rev. B \textbf{75}, 125314 (2007).
\bibitem{chan11} C.-K. Chan and L. J. Sham, Phys. Rev. A \textbf{84}, 032116 (2011).
\bibitem{chan12} C.-K. Chan and L. J. Sham, J. Opt. Soc. Am. B \textbf{29}, A25, (2012).

\bibitem{gessner11} M. Gessner and H.-P. Breuer, Phys. Rev. Lett. \textbf{107}, 180402 (2011).
\bibitem{roghani11} M. Roghani, H. Helm, and H.-P. Breuer, Phys. Rev. Lett. \textbf{106}, 040502 (2011).
\bibitem{laine12} E.-M. Laine, H.-P. Breuer, J. Piilo, C.-F. Li, and G.-C. Guo, Phys. Rev. Lett. \textbf{108}, 210402 (2012).

\bibitem{lidar12} D. A. Lidar, e-print arXiv:1208.5791.





\end{thebibliography}
\end{document}